\documentclass[Physsubmission, Phys]{SciPost}

\hypersetup{
    colorlinks,
    linkcolor={red!50!black},
    citecolor={blue!50!black},
    urlcolor={blue!80!black}
}

\usepackage[bitstream-charter]{mathdesign}
\DeclareSymbolFont{usualmathcal}{OMS}{cmsy}{m}{n}
\DeclareSymbolFontAlphabet{\mathcal}{usualmathcal}

\newcommand{\bq}{\begin{eqnarray}}
\newcommand{\eq}{\end{eqnarray}}
\newcommand{\eps}{\varepsilon}

\begin{document}

\begin{center}{\Large \textbf{
Feynman integrals for binary systems of black holes \\
}}\end{center}

\begin{center}
Philipp Alexander Kreer\textsuperscript{1},
Robert Runkel\textsuperscript{2} 
and
Stefan Weinzierl\textsuperscript{2$\star$}
\end{center}

\begin{center}
{\bf 1} Physik Department, James-Frank-Stra{\ss}e 1, Technische Universit\"at M\"unchen,
D - 85748 Garching, Germany
\\
{\bf 2} PRISMA Cluster of Excellence, Institut f{\"u}r Physik,
Johannes Gutenberg-Universit{\"a}t Mainz,
D - 55099 Mainz, Germany
\\
* weinzierl@uni-mainz.de
\end{center}

\begin{center}
\today
\end{center}


\definecolor{palegray}{gray}{0.95}
\begin{center}
\colorbox{palegray}{
  \begin{tabular}{rr}
  \begin{minipage}{0.1\textwidth}
    \includegraphics[width=35mm]{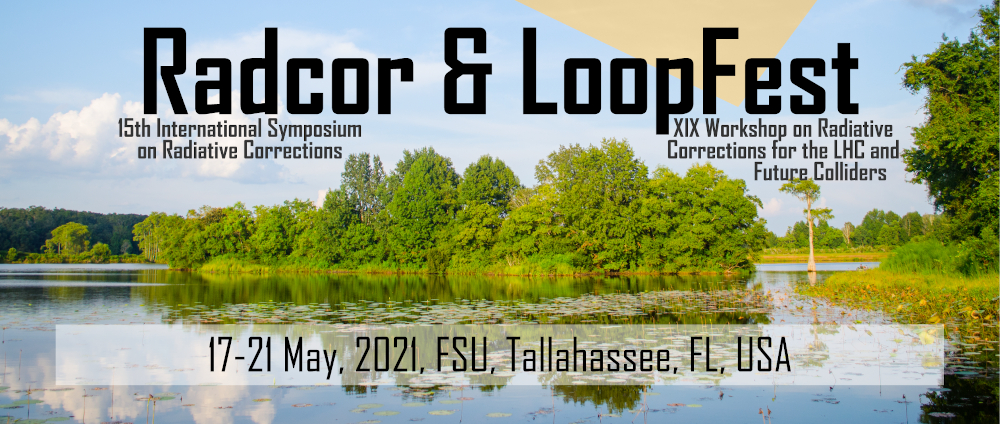}
  \end{minipage}
  &
  \begin{minipage}{0.85\textwidth}
    \begin{center}
    {\it 15th International Symposium on Radiative Corrections: \\Applications of Quantum Field Theory to Phenomenology,}\\
    {\it FSU, Tallahasse, FL, USA, 17-21 May 2021} \\
    \doi{10.21468/SciPostPhysProc.?}\\
    \end{center}
  \end{minipage}
\end{tabular}
}
\end{center}

\section*{Abstract}
{\bf
The initial phase of the inspiral process of a binary black-hole system
can be described by perturbation theory.
At the third post-Minkowskian order a two-loop double box graph, known
as H-graph, contributes. 
In this talk we report how all master integrals of the H-graph with equal masses can be expressed 
up to weight four in terms of multiple polylogarithms.
We also discuss techniques for the unequal mass case.
The essential complication (and the focus of the talk) is the occurrence of several square roots.
}



\section{Introduction}
\label{sect:introduction}

The initial phase of the inspiral process
of a binary system producing gravitational waves
can be described by perturbation theory \cite{Buonanno:1998gg,Buonanno:2000ef,Damour:2012mv,Damour:2017ced,Bini:2020nsb,Bini:2020hmy}.
Effective field theory methods provide a link between general relativity and particle physics \cite{Goldberger:2004jt,Cheung:2018wkq,Porto:2016pyg,Levi:2018nxp} and there has been a fruitful interplay between these fields in recent years \cite{Foffa:2016rgu,Foffa:2019hrb,Bini:2020uiq,Bini:2020rzn,Bjerrum-Bohr:2018xdl,Cristofoli:2019neg,Kosower:2018adc,Bern:2019nnu,Bern:2019crd,Bern:2021dqo,Blumlein:2019zku,Blumlein:2019bqq,Blumlein:2020pog,Blumlein:2020znm,Blumlein:2020pyo,Blumlein:2021txj,Foffa:2019rdf,Foffa:2019yfl,Kalin:2019rwq,Kalin:2019inp,Kalin:2020mvi,Kalin:2020fhe,Liu:2021zxr,Herrmann:2021lqe,DiVecchia:2021bdo,Bjerrum-Bohr:2021vuf}.
Two expansions are important:
The post-Newtonian expansion is an expansion in the weak gravitational field limit and the small velocity limit.
The post-Minkowskian expansion is an expansion in the weak gravitational field limit.

\begin{figure}[h]
\centering
\includegraphics[scale=1.0]{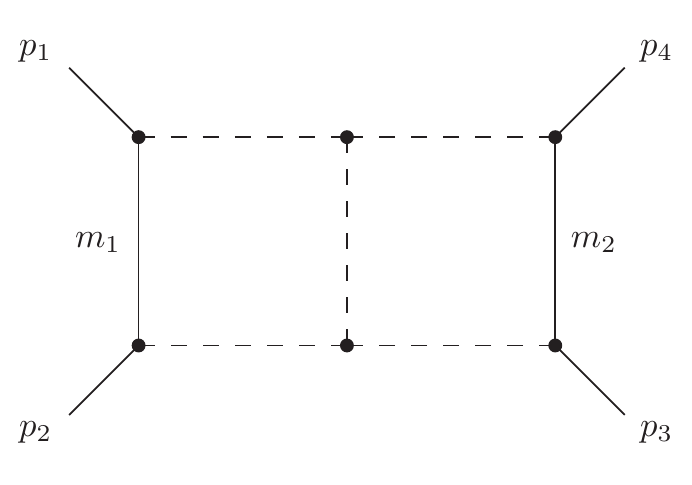}
\caption{The H-graph. Solid lines denote massive objects, dashed lines denote massless particles.}
\label{fig_H-graph}
\end{figure}

Let us focus on the third order of the post-Minkowskian expansion of a binary system.
The most complicated graph entering at this order is shown in fig.~\ref{fig_H-graph} and is called the H-graph.
The name H-graph stems from the fact that the gravitons form the letter ``H'' 
(which in our figure is rotated by $90^\circ$).
In order to describe the kinematics we introduce the Mandelstam variables
\bq
 s \; = \; \left(p_1+p_2\right)^2,
 & &
 t \; = \; \left(p_2 + p_3\right)^2.
\eq
The external momenta are on-shell:
\bq
 p_1^2 \; = \; p_2^2 \; = \; m_1^2,
 & &
 p_3^2 \; = \; p_4^2 \; = \; m_2^2.
\eq
For a binary system the limit $|s| \ll t, m_1^2, m_2^2$ is relevant.
Nevertheless, a calculation without approximations is helpful.
We distinguish two cases:
\begin{alignat}{2}
 & \mbox{Equal mass case}: & & m_1=m_2=m. \nonumber \\
 & \mbox{Unequal mass case}: & \hspace*{2mm} & m_1 \neq m_2. 
\end{alignat}
The equal mass case has been first studied in \cite{Bianchi:2016yiq}.
The first steps of our calculation are standard:
We use integration-by-parts identities \cite{Tkachov:1981wb,Chetyrkin:1981qh}
to derive the differential equation for the master integrals $I$
\cite{Kotikov:1990kg,Kotikov:1991pm,Remiddi:1997ny,Gehrmann:1999as}
\bq 
\label{dgl}
 d I & = & A I.
\eq
The differential equation can be cast to an $\eps$-form \cite{Henn:2013pwa}:
\bq
 A & = & \eps \sum\limits_{k=1}^{N_{\mathrm{letter}}} C_k \omega_k,
 \;\;\;\;\;\;
 \omega_k \; = \; d \ln f_k.
\eq
In our case the $f_k$'s are algebraic functions of the kinematic variables, i.e. 
the $f_k$'s may contain square roots.
Our system of master integrals is characterised by the number kinematic variables
$N_{\mathrm{kin}}$, the number of master integrals $N_{\mathrm{master}}$, the number of distinct square roots $N_{\mathrm{root}}$ and
the number of letters $N_{\mathrm{letter}}$.
Table~\ref{table_essential_numbers} summarises these numbers for the equal and unequal mass case.
\begin{table}[h]
\centering
\begin{tabular}{|l|l|}
 \hline
 Equal mass case: & Unequal mass case: \\
 \hline 
 $2$ kinematic variables $s/m^2$, $t/m^2$ & $3$ kinematic variables $s/m_1^2$, $t/m_1^2$, $m_2^2/m_1^2$ \\
 $25$ master integrals & $40$ master integrals \\
 $4$ square roots & $6$ square roots \\
 $17$ dlog-forms & $29$ dlog-forms \\
 \hline
\end{tabular}
\caption{The essential numbers for the equal and unequal mass case.}
\label{table_essential_numbers}
\end{table}
The differential equation~(\ref{dgl}) is solved in terms of Chen's iterated integrals $I_\gamma(\omega_{i_1},\dots,\omega_{i_r})$ \cite{Chen}.
We are interested in the question if these iterated integrals (say up to weight four) 
can be converted to standard functions (i.e. multiple polylogarithms).
The complication is given by the occurrence of the square roots.
In the unequal mass case the occurring square roots are
\begin{align}
 r_1
 & = 
 \sqrt{-s \left(4 m_1^2-s\right)},
 &
 r_4
 & = 
 \sqrt{-s \left[4 m_1^2 m_2^4 - s \left( m_1^2-t\right)^2\right]},
 \\
 r_2
 & = 
 \sqrt{-s \left(4 m_2^2-s\right)},
 &
 r_5
 & = 
 \sqrt{-s \left[4 m_1^4 m_2^2 -s \left( m_2^2 -t \right)^2\right]},
 \nonumber \\
 r_3 
 & =   
 \sqrt{\left[\left(m_1-m_2\right)^2-t\right]\left[\left(m_1+m_2\right)^2-t\right]},
 &
 r_6 
 & =   
 \sqrt{\left[\left(m_1-m_2\right)^2-s-t\right]\left[\left(m_1+m_2\right)^2-s-t\right]}.
 \nonumber 
\end{align}
In the equal mass case $r_1$ and $r_2$ become identical, and so do $r_4$ and $r_5$.

\begin{figure}[h]
\centering
\begin{tabular}{lccc}
 &
\includegraphics[scale=0.4]{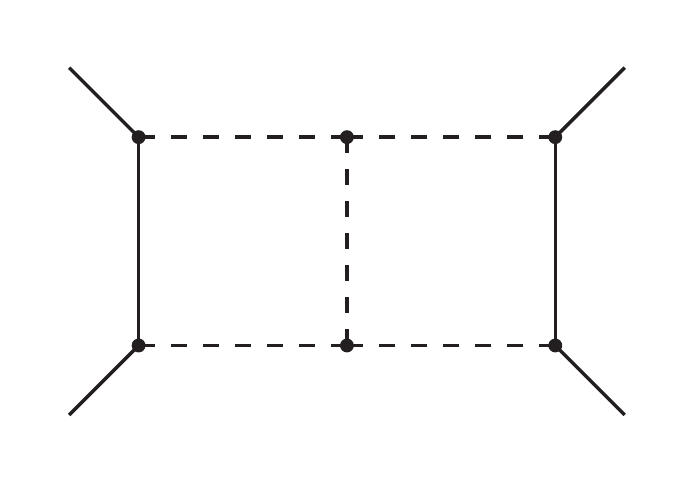} &
\includegraphics[scale=0.4]{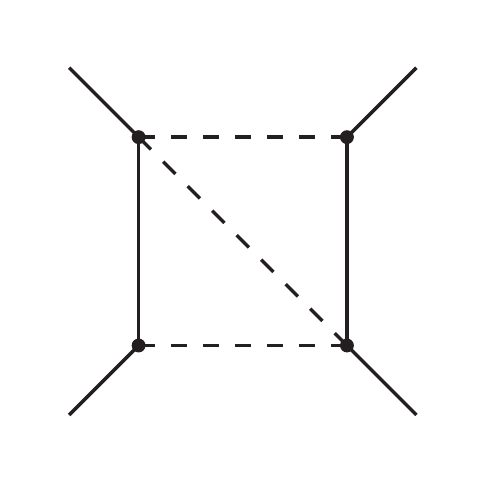} &
\includegraphics[scale=0.4]{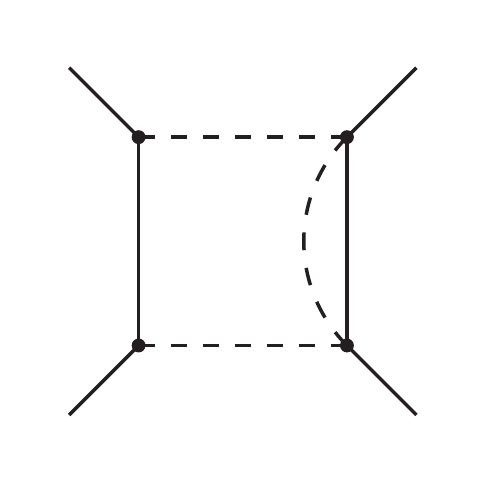} \\
 unequal: & 4 masters & 5 masters & 3 masters \\
 equal: & 3 masters & 4 masters & 3 masters \\
\end{tabular}
\caption{Topologies with a dependence up to weight $4$ on a non-trivial subset of the square roots.}
\label{fig_H-graph_subtopologies}
\end{figure}

We remark that 
the scalar double box integral with no dots and no irreducible scalar products 
has up to weight four a rather simple expression in terms of multiple polylogarithms.
In the equal mass case we have
\bq
 \frac{1}{\eps^4 s^2 r_3} I_{1111111}
 & = &  
 4 \left[ G\left(1,1,1;\bar{y}\right) + \zeta_2 G\left(1;\bar{y}\right) \right] \eps^3
 + 4 \left\{
           2 G\left(2,1,1,1;\bar{y}\right)
 \right.
 \nonumber \\
 & &
 \left.
           + 2 G\left(0,1,1,1;\bar{y}\right)
           - 2 G\left(1,1,2,1;\bar{y}\right)
           - 2 G\left(1,1,0,1;\bar{y}\right)
 \right. \nonumber \\
 & & \left.
           + 2 \zeta_2 \left[ G\left(2,1;\bar{y}\right) + G\left(0,1;\bar{y}\right) - G\left(1,1;\bar{y}\right) \right]
           - \zeta_3 G\left(1;\bar{y}\right) 
 \right. \nonumber \\
 & & \left.
           + 2 \left[ G\left(1,1,1;\bar{y}\right) + \zeta_2 G\left(1;\bar{y}\right) \right] \left[ G\left(1,\bar{x}\right) -2 \ln\left(\bar{x}\right) \right] 
   \right\} \eps^4
 + {\mathcal O}\left(\eps^5\right),
 \nonumber \\
 & &
 s \; = \; - \frac{\bar{x}^2}{1-\bar{x}} m^2,
 \;\;\;\;\;\;
 t \; = \; - \frac{\bar{y}^2}{1-\bar{y}} m^2.
\eq
and similar for the unequal mass case.
We are interested in all master integrals up to weight four.
The remaining master integrals in the top sector as well as 
master integrals from sub-sectors will involve non-trivial subsets of the square roots up to weight $4$.
The most challenging topologies are shown in fig.~\ref{fig_H-graph_subtopologies}.


\section{Rationalisation of square roots}
\label{sect:rationalisation}

By a change of variables it is sometimes possible to rationalise a square root.
Let us look at a simple example.
Consider a dlog-form with a square root:
\bq
 \omega
 & = & 
 d \ln\left(\frac{2m_1^2-s-r_1}{2m_1^2-s+r_1}\right)
 \; = \; \frac{2s}{r_1} \left( \frac{ds}{s} - \frac{dm_1^2}{m_1^2} \right)
\eq
The transformation
\bq
 - \frac{s}{m_1^2} & = & \frac{\left(1-x\right)^2}{x}
\eq
rationalises the square root:
\bq
 \omega
 & = &
 2 d\ln\left( x \right)
 \; = \; 2\frac{dx}{x}
\eq
Let us assume that the differential equation is in $\eps$-dlog-form, where the arguments of the logarithms
contain square roots.
There are algorithms to rationalise square roots \cite{Besier:2018jen,Besier:2019kco}.
If we can simultaneously rationalise all square roots, all integrals can be expressed
in terms of multiple polylogarithms.
It can be hard to prove that the set of square roots cannot be rationalised simultaneously \cite{Besier:2019hqd}.
Even if one can prove that a set of square roots cannot be rationalised simultaneously,
this does not imply that the Feynman integrals cannot be expressed in terms of multiple polylogarithms \cite{Heller:2019gkq}.
Hence, being able to rationalise simultaneously all square roots is a sufficient condition that the result can be expressed in
terms of multiple polylogarithms, but not a necessary condition.
If a set of square roots cannot be rationalised simultaneously, other methods like symbol calculus \cite{Heller:2019gkq,Heller:2021gun}
or introducing additional variables \cite{Papadopoulos:2014lla,Papadopoulos:2014hla,Papadopoulos:2015jft,Canko:2020ylt} may express the Feynman integrals in terms of multiple polylogarithms.

Multiple square roots appear not only in the Feynman integrals associated to the H-graph,
but also in Feynman integrals associated to other processes.
Example are Bhabha scattering \cite{Henn:2013woa} or Drell-Yan \cite{Bonciani:2016ypc}.
As we study more two-loop integrals with several kinematic variables, multiple square roots
become more frequent.
We want to learn how to handle them.

We therefore study the case of the H-graph in more detail.
We first note that 
the last square root $r_6$ appears up to weight $4$ only in one master integral (one master integral from the sub-topology shown in the middle in fig.~\ref{fig_H-graph_subtopologies}).
This master integral can be computed in the Feynman parameter representation and 
evaluates to multiple polylogarithms following the lines of \cite{Brown:2008,Panzer:2014caa,Heller:2019gkq}.
This holds in the equal mass case and in the unequal mass case.

In the equal mass case we then have the following situation:
The remaining 24 master integrals involve up to weight $4$ only three square roots.
These square roots can be rationalised simultaneously and therefore
all master integrals evaluate up to weight $4$ to multiple polylogarithms \cite{Kreer:2021sdt}.

\begin{figure}[h]
\centering
\includegraphics[scale=1.0]{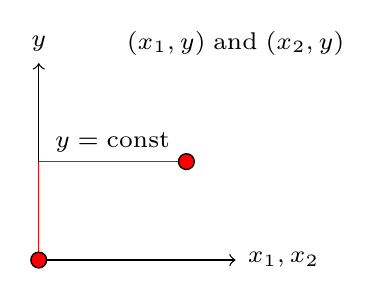}
\hspace*{5mm}
\includegraphics[scale=1.0]{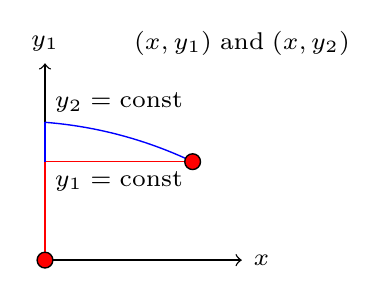}
\caption{The left picture shows two parametrisations of the same integration path, the right picture 
shows two different integration paths.}
\label{fig_integration_path}
\end{figure}

The unequal mass case is more complicated:
Each master integral is a linear combination of iterated integrals.
Each iterated integral of the master integrals contains up to weight $4$ 
no more than $3$ distinct roots.
In the remaining 39 master integrals up to weight $4$ 
each occurring triple of distinct roots can be rationalised
simultaneously.
In other words: we can rationalise simultaneously any occurring triple $(r_i,r_j,r_k)$ 
from $\{r_1,r_2,r_3,r_4,r_5\}$, but we may have to use different rationalisations
for different triples.
This raises the question whether we are allowed to use different rationalisations for different iterated integrals.
We have to distinguish two cases:
\begin{enumerate}
\item Different rationalisations may correspond to different parametrisations of the same integration path (left picture of fig.~\ref{fig_integration_path}).
\item Different rationalisations may correspond to different integration paths (right picture of fig.~\ref{fig_integration_path}).
\end{enumerate}
Case $1$ is unproblematic and we therefore discuss case $2$.
A single iterated integral $I_\gamma(\omega_1,\dots,\omega_r)$ is in general path dependent.
The linear combination of iterated integrals in the 
$\eps^j$-term of the $i$-th master integral $I_i^{(j)}$ is path independent,
this is ensured by the integrability condition of the differential equation $dA-A\wedge A=0$.
This allows us to use different integration paths for $I_{i_1}^{(j_1)}$ and $I_{i_2}^{(j_2)}$.

Now let's look at a single expression $I_i^{(j)}$:
We would like to split
\bq
 I_i^{(j)}
 & = &
 I_{i,a}^{(j)} + I_{i,b}^{(j)}
\eq
and use different integration paths for $I_{i,a}^{(j)}$ and $I_{i,b}^{(j)}$.
This is only allowed if $I_{i,a}^{(j)}$ and $I_{i,b}^{(j)}$ are path independent.
In order to give a criteria when a linear combination of iterated integrals is path independent we first introduce
the bar notation for the tensor algebra (we assume that all $\omega_j$'s are closed):
\bq
 \left[ \omega_1 | \omega_2 | \dots | \omega_r \right]
 & = &
 \omega_1 \otimes \omega_2 \otimes \dots \otimes \omega_r,
 \nonumber \\
 d \left[ \omega_1 | \omega_2 | \dots | \omega_r \right]
 & = &
 \sum\limits_{j=1}^{r-1} \left[ \omega_1 | \dots | \omega_{j-1} | \omega_j \wedge \omega_{j+1} | \omega_{j+2} | \dots | \omega_r \right].
\eq
We associate to a linear combination of iterated integrals
\bq
 I
 \; = \;
 \sum\limits_{j=1}^r \sum\limits_{i_1,\dots,i_j}
 c_{i_1 \dots i_j} I_\gamma\left(\omega_{i_1},\dots,\omega_{i_j}\right)
 & \Rightarrow &
 B
 \; = \;
 \sum\limits_{j=1}^r \sum\limits_{i_1,\dots,i_j}
 c_{i_1 \dots i_j} \left[\omega_{i_1}|\dots|\omega_{i_j}\right]
\eq
By a theorem of Chen \cite{Chen}
the linear combination $I$ is path independent if and only if $dB=0$.

\begin{figure}[h]
\centering
\includegraphics[scale=1.0]{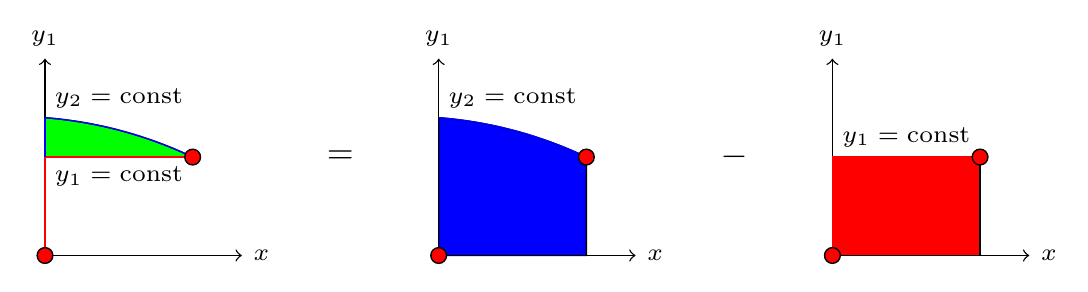}
\caption{The difference of two integration paths defines an area (green), which we write may write as the difference of two areas, where
each area only knows about one integration path.}
\label{fig_stokes}
\end{figure}

For the case at hand, we observe that $\omega_i \wedge \omega_j$ may involve less square roots than $\omega_i \otimes \omega_j$.
Although $I_{i,a}^{(j)}$ and $I_{i,b}^{(j)}$ may be path dependent, we may find $I_{\mathrm{subtr}}^{(j)}$
compatible with two rationalisations such that
\bq
 \left(I_{i,a}^{(j)}-I_{\mathrm{subtr}}^{(j)}\right)
 & \mbox{and} &
 \left(I_{i,b}^{(j)}+I_{\mathrm{subtr}}^{(j)}\right)
\eq
are path independent. 
We may then evaluate $(I_{i,a}^{(j)}-I_{\mathrm{subtr}}^{(j)})$ with a rationalisation corresponding to an integration path 
$\gamma_a$ and $(I_{i,b}^{(j)}+I_{\mathrm{subtr}}^{(j)})$ with a rationalisation corresponding to an integration path $\gamma_b$.
The subtraction terms can be obtained from Stokes' theorem by integrating $\omega_i \wedge \omega_j$,
as sketched in fig.~\ref{fig_stokes}.
We remark that in mathematical terms, $I_{\mathrm{subtr}}^{(j)}$ is related to Massey products.

Let us discuss another pitfall, which may occur when using different rationalisations.
This pitfall may occur already for different parametrisations of the same integration path
and is related to trailing zeros.
Consider the transformation $x = 2 x'$ and the dlog-form
\bq
 \omega_0 \; = \; d\ln(x)
 \; = \;
 d\ln(2x')
 \; = \;
 d \left[ \ln(2) + \ln(x') \right]
 \; = \;
 d\ln(x').
\eq
Consider the apparent contradiction
\bq
\label{contradiction}
 \int\limits_0^{x_f} \omega_0 
 \; = \; 
 \ln(x_f)
 & \neq &
 \ln(x_f) - ln(2) 
 \; = \; 
 \ln(x_f') 
 \; = \; 
 \int\limits_0^{x_f'} \omega_0.
\eq
The solution to this contradiction is as follows:
The two integrals appearing in eq.~(\ref{contradiction}) are divergent integrals.
Although it is common practice to define
\bq
 \int\limits_0^{x_f} \frac{dx}{x} & = & \ln(x_f),
\eq
we should not forget what this notation actually means:
We first introduce a lower cut-off $\lambda$ as a regulator.
In a second step we employ a ``renormalisation scheme'' and remove all $\ln(\lambda)$-terms.
It is now clear what the solution has to be:
A transformation $x=2x'$ induces a change of the ``renormalisation scheme''.
In practice this boils down to that we isolate all trailing zeros in $x'$ and substitute
\bq
 \ln(x') & \rightarrow & \ln(x') + \ln(2).
\eq

\section{Conclusion}

In this talk we discussed the computation of the master integrals of the two-loop H-graph, relevant to gravitational waves.
The differential equation for the master integrals involves four square roots in the equal mass case and six square roots
in the unequal mass case.
The major challenge is a method for an efficient computation for the case 
where not all square roots can be rationalised simultaneously.
In the equal mass case we found that all master integrals up to weight 4 can be expressed in terms of multiple polylogarithms.
For the unequal mass case we reported on techniques to render sub-expressions of iterated integrals path-independent. 

{\footnotesize
\bibliography{/home/stefanw/notes/biblio}
}

\nolinenumbers

\end{document}